# Predicting Effective Diffusivity of Porous Media from Images by Deep Learning

Haiyi Wu,[1] Wen-Zhen Fang,[1,2] Qinjun Kang,[3,*] Wen-Quan Tao,[2,*] and Rui Qiao[1,*]

[1] Department of Mechanical Engineering, Virginia Tech, Blacksburg, VA 24061, USA

[2] Key Laboratory of Thermo-Fluid Science and Engineering, MOE, Xi'an Jiaotong University, Xi'an, China

[3] Earth and Environmental Sciences Division, Los Alamos National Laboratory, Los Alamos, NM 87545, USA

**Abstract.** We report the application of machine learning methods for predicting the effective diffusivity ($D_e$) of two-dimensional porous media from images of their structures. Pore structures are built using reconstruction methods and represented as images, and their effective diffusivity is computed by lattice Boltzmann (LBM) simulations. The datasets thus generated are used to train convolutional neural network (CNN) models and evaluate their performance. The trained model predicts the effective diffusivity of porous structures with computational cost orders of magnitude lower than LBM simulations. The optimized model performs well on porous media with realistic topology, large variation of porosity (0.28-0.98), and effective diffusivity spanning more than one order of magnitude ($0.1 \lesssim D_e < 1$), e.g., >95% of predicted $D_e$ have truncated relative error of <10% when the true $D_e$ is larger than 0.2. The CNN model provides better prediction than the empirical Bruggeman equation, especially for porous structure with small diffusivity. The relative error of CNN predictions, however, is rather high for structures with $D_e < 0.1$. To address this issue, the porosity of porous structures is encoded directly into the neural network but the performance is enhanced marginally. Further improvement, i.e., 70% of the CNN predictions for structures with true $D_e < 0.1$ have relative error <30%, is achieved by removing trapped regions and dead-end pathways using a simple algorithm. These results suggest that deep learning augmented by field knowledge can be a powerful technique for predicting the transport properties of porous media. Directions for future research of machine learning in porous media are discussed based on detailed analysis of the performance of CNN models in the present work.

* Correspondence and requests for materials should be addressed to R.Q. (ruiqiao@vt.edu), Q.K. ( qkang@lanl.gov), and W.-Q.T. (wqtao@mail.xjtu.edu.cn).

# I. Introduction

Predicting the transport properties of porous materials from their structure is important in numerical simulations of a wide range of engineering problems, e.g., extraction of nature gas from shale reservoirs,[1-4] drying of building materials,[5] and charging/discharging of Li-air batteries. The transport properties of interest include effective diffusivity, permeability, thermal conductivity, among others.[1] A classical approach for calculating these effective transport properties is the pore scale simulations, in which the governing equations for the related transport phenomena are solved within the porous media.[2,6,7] While this approach can be highly accurate, its computational cost is significant for porous media with large dimension and/or small pore sizes. In fact, for porous media that undergo long operation during which their structure evolves (e.g., the pores in electrodes of Li-air batteries are gradually clogged during discharging), this approach can be prohibitively expensive because their effective transport properties may need to be evaluated for millions of times. As such, the effective transport properties of porous media are often computed using empirical correlations or effective medium theories with their structure information (e.g., porosity) as input. Such an approach needs little computational cost and can be very accurate for some specific (often idealized) classes of porous media. However, because typically only a few structure parameters of the porous media are used as input in this approach, its prediction for complex porous media often lacks specificity and can be inaccurate. Indeed, it remains a great challenge to develop methods for predicting the effective transport properties of porous media that require low computational cost but offer high accuracy for diverse porous structures.

Machine learning can potentially be an effective approach for tackling the above challenge. Deep neural networks have demonstrated good predictive power when their input and output have important correlation with each other. Furthermore, image-based learning has been shown to be able to extract important physical features from images.[8,9] Because the effective transport properties (in particular, the effective diffusivity) of porous media is largely determined by their structure which can be conveniently represented using their binary images, conceivably, one can develop a surrogate deep learning model to extract key geometrical features from images of porous media and predict their transport properties. In terms of implementation, the application of deep neural network typically requires a training dataset, which can be generated numerically or experimentally for porous media. Next, a training model is constructed and trained using the dataset. Finally, the trained model can be used to predict the effective transport properties of new porous structures without repeating the training process. This general strategy resembles the investigation of image classification, where images are taken as the input and trained deep learning models predict the classification label (e.g. "dog" or "cat") of images by identifying their key features.[10-12] As demonstrated in the studies of image classification, typically, a trained model can be used to make predictions with low computational cost. Therefore, a deep learning model, if well-constructed and trained, can potentially predict the effective transport properties of porous media both accurately and efficiently.

Of the many deep neural network models, convolutional neural network (CNN) [13] is commonly applied to analyze visual imagery and has achieved much success in image classification. Recently, CNN has also been adopted to study the effective properties of complex materials and showed much potential for efficient and accurate prediction of a material's effective properties from its structure (e.g., presented in the form of images). For example, researchers have used CNN to predict the effective permeability and stiffness of materials from their microstructures.[8,9,14-16] In particular, three dimensional CNN has shown to capture the nonlinear mapping between material microstructure and its effective stiffness [8]. Study of the prediction of permeability from images of porous media using CNN has provided useful insights in understanding the correlation between geometric features and transport properties.[14,16] The features of connectivity between nearby pixels in the image of a porous structure can be extracted by performing convolution with many possible cross shape templates. It has been shown that the features thus extracted can be used to make better prediction of permeability than using the geometric measurements (Minkowski functionals).[16] Furthermore, it has been pointed out that deep learning approach can be further improved by incorporating physical parameters of porous media that are known to affect effective permeability.[14] Nevertheless, the porous structure in previous works is relatively well-defined and of limited geometrical variablity, and some of the chanllenging topologies (i.e. trapped and dead-end pathways) that are commonly seen in porous media were not included. Consequently, it is still not clear whether the CNN model can accurately predict the effective transport properties of practical porous media with diverse geometries and challenging topologies. Delineating the impact of such diversity and complexity of porous media on the performance of CNN models and building more sophisticated deep learning models to deal with them are important for the practical application of CNN models in porous media research.

In this work, we establish a computational framework to predict the effective diffusivity of porous media from their images using CNN. We focus on the effective diffusivity because of its importance in practical applications of porous media. The proposed framework is demonstrated in porous media with a wide range of porosity ($0.28 - 0.98$) and diverse/complex structures (e.g., tortuous diffusion pathways and trapped and dead-end regions). Different filter sizes and convolutional layers are tested during the cross-validation step in order to select the best hyperparameters for CNN. Physical parameter (the porosity) is also combined with the extracted features by CNN with different weights to generate better predictions. The CNN can predict the effective diffusivity of most of the porous media with less than 10% truncated relative error. Nevertheless, the prediction error, especially the relative error, increases as the true diffusivity of the porous media becomes smaller (especially if $D_e$ <0.1). The large error is attributed to the complex transport behavior in porous media with low diffusivity, where the porous structure can be highly heterogeneous with highly tortuous diffusion pathways and numerous trapped areas or dead-end paths.

The rest of the manuscript is structured as follows: In Section II, we detail the computational framework and methods used for predicting the effective diffusivity of porous media by deep learning models. In Section III, the performance of the developed model is examined and discussed. We first quantify the performance of CNN and compare the CNN prediction with the empirical Bruggeman equation predictions. Then, we attempt to improve the regular CNN prediction by leveraging field knowledge in the development and application of CNN models. Finally, conclusions and discussion on the future research directions of machine learning for porous media research are presented in Sec. IV.

## II. Computational framework

In this section, we discuss the deep learning model for predicting the effective diffusivity of porous media with diverse and challenging structures. In part A, we present the methods for generating the dataset for deep learning. In part B, we present the architecture of our CNN model and summarize the computational framework of using CNN to predict the effective diffusivity of porous media from their images. Finally, the methods for training and testing of the CNN models are presented in part C.

## A. Generation of datasets

The dataset for training, validating, and testing our deep learning models includes the structure of porous media (in the form of images) and their corresponding effective diffusivity. Without losing generality, we will focus on two-dimensional (2D) porous media with a square shape. The microstructures of 2D porous material are generated using the quartet structure generation set (QSGS) method, a popular method in the porous media field.[17] Detailed descriptions of this method can be found in the literature and we only outline its key steps:[17] (1) The computational domain is partitioned into square cells. (2) Solid "seeds" are randomly distributed in the domain based on a distribution probability, $c_d$, which is smaller than the target porosity of the porous media. This is accomplished by assigning a random number to each cell and the cells whose assigned random number is less than $c_d$ are selected as the "seeds". (3) Grow the "seeds" to their neighboring cells based on the directional growth probability, $P_i$. To this end, a random number is assigned to each of the neighboring cell of a solid seed. If the random number of a neighboring cell is less than $P_i$, it will become part of the growing solids. (4) Repeat steps (2) and (3) until the target porosity is reached in the domain. Together, the above steps produce binary images of porous media. In these images, individual pores or grains are fully resolved, and each pixel is either a pore or solid node and is denoted with a binary value of 0 (pore space) or 1 (solid phase).

Using the above method, 2-D porous media are generated within a $200 \times 200$ (pixel) area. This size of the porous structure is chosen so that (1) the microstructures are large enough to capture the range of the topologies and transport behavior occurring in realistic porous media and (2) the porous structures are small enough so that a large number of porous structures and their effectivity diffusivity can be obtained at a reasonable computational cost. To ensure that the dataset include a wide variety of porous structures, 1960 samples are generated with porosity ($\varepsilon$) of 0.28, 0.29, ..., 0.98. For each porosity, 28 samples are generated. Figure 1a shows the representative images of the porous samples generated. A wide variety of structures featuring tortuous transport pathways, trapped regions, and dead-end pores are obtained in samples with porosity smaller than 0.5.

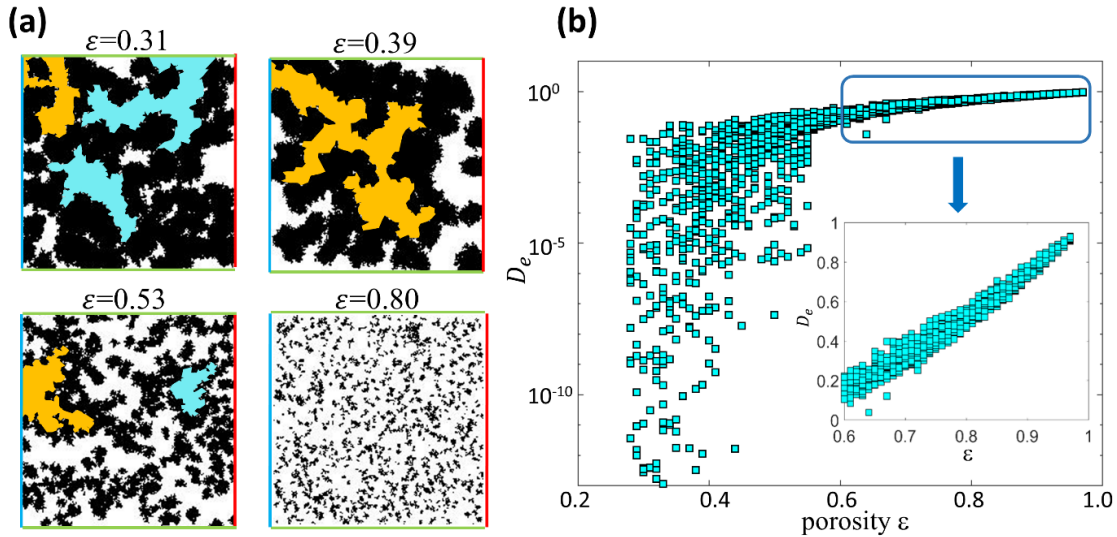

**Figure 1**. (**a**) Representative images of the two-dimensional porous media generated for the deep learning model. The white and black color denote the pore space and solid phase, respectively. The inlet (outlet) for the mass diffusion through the porous media are marked using blue (red) lines. The green lines at the top and bottom side of the porous media denote the periodic boundary in the LBM simulation. The orange areas represent the dead-end pathways (i.e., the pathways that are connected only to the inlet or outlet of the porous structure) and the cyan areas denote the trapped pore space (i.e., isolated pore space that is not connected to the porous structure's inlet and outlet). (**b**) The distribution of the effective diffusivity of porous media generated in this work.

We next compute the effective diffusivity of the porous structures generated above. The molecular transport within the porous structure is assumed to follow Fick's Law with a constant diffusion coefficient $\widetilde{D}_0$. Therefore, the molecular diffusion within the porous structure obeys the Laplace equation with a zero-flux boundary condition on pore surfaces. To compute the effective diffusivity of each porous structure, a uniform concentration difference ($\Delta\widetilde{C}$) is imposed between the left and right boundaries of each porous structure and the periodic boundary condition is imposed on the top and bottom boundaries, respectively. The Laplace equation is solved using the lattice Boltzmann method (LBM). Specifically, a two-dimensional, nine-velocity (D2Q9) LB model is adopted to simulate the diffusion process inside the porous structures. Different from traditional numerical methods by discretizing the Laplace equation in the pore space, LBM solves the evolution equation of the concentration distribution functions

$$g_i(x + e_i\Delta t, t + \Delta t) - g_i(x, t) = -\frac{1}{\tau}(g_i(x,t) - g_i^{eq}(x,t)) \tag{1}$$

where $g_i$ is the concentration distribution function at the space location $x$ and time $t$ along $i$ direction; $\tau$ is the relaxation time coefficient; $g_i^{eq} = \omega_i C$ is the corresponding equilibrium concentration distribution function, where $C = \sum g_i$ is the macroscopic local concentration, $\omega_i$ is weight parameter, and $\omega_0 = 4/9$, $\omega_{1-4} = 1/9$, $\omega_{5-8} = 1/36$. In Eq. (1), $e_i$ is the discrete velocity given by

$$e_i = \begin{bmatrix} 0 & 1 & 0 & -1 & 0 & 1 & -1 & -1 & 1 \\ 0 & 1 & 1 & 0 & -1 & 1 & 1 & -1 & -1 \end{bmatrix} \tag{2}$$

The relation between the intrinsic gas diffusion coefficient and the relaxation time coefficient is

$$\widetilde{D}_0 = \frac{1}{3}(\tau - 0.5)\frac{\Delta x^2}{\Delta t} \tag{3}$$

After the diffusion within the porous structure reaches a steady state, the cross-section averaged diffusive flux through the structure is obtained by[18]

$$J_x = \frac{\int_0^{L_y} \sum e_{i,x} g_i \frac{\tau - 0.5}{\tau} dy}{L_y} \tag{4}$$

where $L_y$ is the domain size in the direction normal to the overall diffusion flux. The dimensional effective diffusivity of the porous structure is then determined using Fick's law, i.e., $\widetilde{D}_e = J_x L_x/\Delta C$ ($L_x$ is the domain size in the direction of the overall diffusion flux). Because $\widetilde{D}_e$ is linearly proportional to the molecule diffusion coefficient in the pore space ($\widetilde{D}_0$), the dimensionless effective diffusivity $D_e = \widetilde{D}_e/\widetilde{D}_0$ is used hereafter. The effectivity diffusivity of the porous structures generated above spans ~$10^{-10}$ to 1.0 and is shown as a function of the porosity of the porous structures in Fig. 1b.

The data generated above (porous structures and their $D_e$) are randomly divided into the training dataset (60% of the whole dataset), validation dataset (11.4%) and testing dataset (28.6%). The training dataset is used to optimize the parameters of the CNN model so that the model can describe the training dataset as accurately as possible. The validation dataset is used to select hyperparameters and avoid overfitting the CNN model. The testing dataset is used to evaluate the predictive performance of the trained CNN model.

## B. Convolutional neural network for predicting effective diffusivity

The basic concepts of classical and convolutional neural networks, along with terminologies including hyperparameters and learnable parameters, are reviewed in the *Supporting Information*. These models, especially the CNN, have been used for image classification with great success and is being explored for predicting the effective permeability of porous media. Inspired by these works, we adapt the CNN model for predicting the effective diffusivivity of porous media from their images,[11,19,20] and the architecture of our model is shown in Fig. 2. The binary image of a porous structure, in which a pixel with a value of 1 (0) corresponds to the solid phase (pore space), is the input of the CNN. We note that the binary nature of the input image is consistent with the format of the computationally generated porous media in this work (see Section II.A) and is not a result of the binarization of grayscale images as have been done in many image recognition studies. Because the desired pixel size of images in CNN is $2^m$ ($m$ is an integer), the images of porous structures made in Section A are downsampled to $128 \times 128$ pixels using kernel[21] before feeding to the CNN. Below we outline the different layers in the CNN. We focus on identifying the data flow as well as learnable parameters and their dimensions, but omit numerical implementation details as they are widely available in the literature.

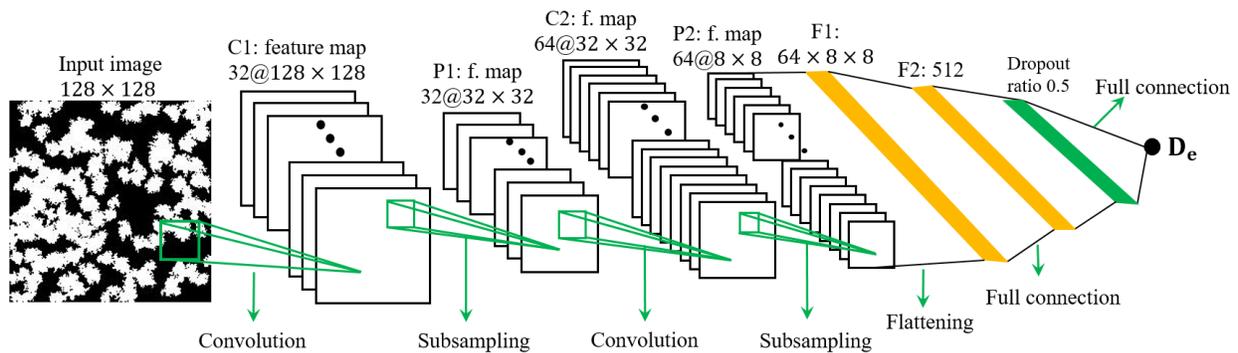

**Figure 2**. The architecture of our regular convolutional neural network (CNN) for predicting the effective diffusivity of 2D porous media from their images.

In this work, we adapt the CNN model have similar archietcture of the AlexNet,[11] Our CNN model has $M$ pairs of convolutional and pooling layers and $P$ fully-connected layers ($M$ and $P$ are both set to 2 in Fig. 2 for illustration purpose). When CNN is used in image-related studies as in this work, the ouput volume of a convolutional or pooling layer is typically termed *feature maps* as the purpose of these layers is to extract features from their input volume.[11, 20] For simplicity, the width of any input/ouput volume of convolutional/pooling layer is always equal to its height in this work. The number of slices of an ouput volume of a convolutional layer ($d_o$) is its depth. To obtain the $\delta$-th slice of the feature map, $\text{F.map}_{\text{conv}}^{\delta}$, a filter is slided over every width and height position of the input volume and the result is the ouput for neuron at each position. By writing $\text{F.map}_{\text{conv}}^{\delta}$ as a $a_o \times a_o$ order-2 tensor ($a_o$ is the width/height of the feature map), this operation can be written as

$$\text{F.map}_{\text{conv}}^{\delta} = \text{ReLU}(W^{\delta} \otimes X_i + B(\delta) \cdot J) \tag{5}$$

where ReLU is the activation function adopted for the neurons in this work.[22] $X_i$ is the input volume of the convolutional layer (for layers C1 and C2 in Fig. 2, $X_i$ are the 2D binary image and the feature map generated by the first pooling layer, respectively). $X_i$ is a $a_i \times a_i \times d_i$ order-3 tensor, where $a_i$ and $d_i$ is the width and depth of the convolutional layer's input volume, respectively. $W^{\delta}$ is the kernel of the $\delta$-th filter of the convolutional layer. It is a $s_{\delta} \times s_{\delta} \times d_i$ order-3 tensor, where $s_{\delta}$ is the filter's spatial extent in the width direction. $W^{\delta} \otimes X_i$ denotes the 2D-convolution of the filter kernel with the input volume, and its expression can be found in numerous textbooks.[23] $B(\delta)$ is the bias for the $\delta$-th filter and $J$ is a $a_o \times a_o$ order-2 tensor with all elements equal to 1.0. In Equation (5), $W^{\delta}$ and $B(\delta)$ ($\delta = 1, 2, \cdots d_o$) must be "learned" during the trainning of CNN and they include a total of $(s_{\delta}^2 \times d_i + 1) \times d_o$ learnable parameters. On the other hand, $a_o$, $s_{\delta}$, $d_o$, and the number of convolutional layers $M$ are hyperparameters to be specified when building the CNN. Generally, the width and depth of the feature map must be large enough to ensure both short- and long-range features can be extracted but small enough to lower comptuational cost and suppress overfitting. In this work, the width of the output feaure maps of a convolutional layer is the same as its input feature map or image. The depth of feature maps geneated by the first and second convolutional layers in our CNN is taken as 32 and 64, respectively.

The feature map generated by each convolutional layer is feed into a pooling layer (see Fig. 2) to obtain a new feature map with the same depth but reduced width. The spatial span, stride, and type of filter used in the pooling layers are also hyperparameters that can be optimized through cross-validation. Here, we use the max pooling with a filter size of $4 \times 4$ and a stride of 4 to downsample the input feature maps.

The feature map generated by the last pooling layer is flatten into a vector ($X_{in,fc1}$; length: $n_{in,fc1}$). This vector is fed to the first fully connected layer with $n_{fc1}$ neurons (see Fig. 2), which produces $F.map_{fc1}$ through Equation (S2) (supporting information) using a $n_{fc1} \times n_{in,fc1}$ matrix $W_{fc1}$ as learnable weights and a vector $B_{fc1}$ (length: $n_{fc1}$) as learnable biases. $n_{fc1}$, a hyperparameter, is taken as 512 in this work. To avoid overfitting, a dropout layer (dropout ratio = 0.5) is applied before the second fully connected layer. The output of the dropout layer is passed to the second fully connected layer to make the final prediction via

$$D_e^{CNN} = W_{fc2}X_{o,drp} + B_{fc2} \tag{6}$$

where $D_e^{CNN}$ is the effective diffusivity predicted by CNN, $X_{o,drp}$ is output of the dropout layer, and $W_{fc2}$ and $B_{fc2}$ are the learnable weight and bias of the second fully connected layer, respectively.

In addition to the CNN above, we also test the deep residual network, Resnet50,[24] a more recent scheme of CNN (see Supporting Information). Because Resnet50 only performs slightly better than the above CNN but its structure is considerably more involved, below we present only the results based on the CNN outlined in Fig. 2. The implementation, results, and comparison of the Resnet50 with the above CNN model are summarized in the *Supporting Information*.

**Using field knowledge to augment CNN model**. A great advantage of CNN is that physical properties can be encoded into the CNN architecture to improve its final prediction. Because the effective diffusivity of a porous structure is strongly correlated with its porosity, following previous work,[14] we also combine the porosity with the flattened feature map of the last pooling layer to form the first fully connected layer in CNN. Specifically, the porosity of each porous structure is added as input to the first fully connected layer of the conventional CNN model with a fixed weight (a variety of weights have been tested with similar results, and a value of 10/4096 is used in the final model here).

Another method that can potentially improve the performance of CNN model is to preprocess the image of porous structures. As shown in Fig. 1a, for structures with low porosity (e.g., $\varepsilon < 0.5$), there exist many trapped areas and/or dead-end paths. These trapped regions are typically larger than the filter size used in the CNN model and thus can be difficult to extract directly using the CNN model. Moreover, the trapped regions are prone to provide faulty information to the feature map and induce noise during the learning process. In this work, before feeding to the CNN, images of porous structures are processed using the 8-connected component analysis[25,26] to eliminate all the trapped pore spaces (see Section III.C.2 for details).

**Summary**. The computation framework for predicting the effective diffusivity of porous structures from their images using CNN models can be summarized as four steps (see Fig. S2):

(a) *Data generation:* Porous media with a wide variety of porosity and pore topologies are generated and their true effective diffusivity $D_e$ (ground truth) is obtained using LBM simulations.

(b) *Augmenting CNN by field knowledge:* Trapped pore spaces are removed in the preprocessing step and physical properties of porous structure (porosity) are combined with feature maps built by the convolutional and pooling layers to serve as the input to the fully-connected layers in CNN.

(c) *Parameterizing CNN.* CNN's hyperparameters and learnable parameters are established through training and cross-validation using the training and validation datasets.

(d) *Deploying CNN.* Using the trained model in step (c) to predict effective $D_e$ of porous media using their images as input.

## C. Parameterization of convolutional neural networks

After the datasets are generated and CNN model is set up, we first need to select all hyperparameters in the CNN model. A CNN model has many hyperparameters such as the choice of the activation function, size of the feature map produced by each convolutional/pooling layer, etc. Although it is desirable to select all parameters through cross-validation, doing so requires prohibitively high computational cost. Therefore, many of these parameters are selected empirically. While this is not a rigorous approach, researches in image classification show that the performance of CNN models is often insensitive to the selection of many hyperparameters. These researches suggest that the number of convolutional layers and the spatial extent (size) of the filters used in these layers usually play the most important role in CNN. Hence, in this work, CNN models with different number of convolutional layers ($M = 2, 3, 4$) and different filter sizes in the convolutional layer ($s \times s = 3x3, 5x5, 7x7$) are evaluated. Other hyperparameters used in the model can be found in Section II.B.2.

We next optimize the weights and biases for all filters in each convolutional and fully connected layer by "learning" the training dataset. The proposed CNN model is implemented in the machine learning framework Tensorflow.[27] The loss function $L$, which quantifies how well the CNN reproduces the training dataset, is defined as the average mean square error over the entire training dataset

$$L = 1/N \sum_{i=1}^{N} \left(D_{e,i}^{LBM} - D_{e,i}^{CNN}\right)^2 \tag{7}$$

The weights of all filters are initialized using a truncated normal distribution with a standard deviation 0.1. All biases are set to 0.1 initially. The training model is converged through minimizing the loss function by the Adam Optimizer[28] with a learning rate $\gamma = 10^{-4}$. The training model stops at 1600 epochs. The optimized weights and biases for all learnable filters at the last epoch are saved in the trained model. The saved model will then be restored for evaluating the testing dataset without redoing the training process.

## III. Results and Discussions

In this section, we first examine how key hyperparameters affect the performance of CNN models using cross-validation. We next present the performance of the regular CNN and compare it with that of a classical empirical model. Then, the performance of field knowledge-informed CNN on a wide range of porous structures is evaluated.

### A. Cross-validation for hyperparameters selection

A trained CNN model often can fit the training dataset used to parameterize it well, but may not perform well on other datasets due to overfitting. To reduce overfitting of the trained model, one usually optimizes hyperparameters of the CNN model through cross-validation. Here, we systematically vary the number of convolutional layers and the filter size in the CNN and cross-validate the trained model using the validation dataset generated in Section II.A. Table 1 summarize the mean square error (MSE) of the effective diffusivity predicted by the CNN model with relative to the true value. We observe that, for a fixed filter size, MSE is smaller for CNN models with fewer convolutional layers. For CNN models with the same number of convolutional layer, the MSE is highest when the filter size is $3 \times 3$, likely because the narrow filters cannot capture some important features spanning moderate to large number of pixels. The MSE is comparable for models with a filter size of $5 \times 5$ and $7 \times 7$. Because the computational cost is higher when filters with $7 \times 7$ size are used, we adopt the CNN models with 2 convolutional layers and a filter size of $5 \times 5$ (see Fig. 2).

Table 1. The mean square error of CNN models with different hyperparameters obtained during cross-validation.

| Number of convolutional layers | Filter size ($s \times s$) | | |
|---|---|---|---|
| | $3 \times 3$ | $5 \times 5$ | $7 \times 7$ |
| 2 layers | $1.38 \times 10^{-4}$ | $8.76 \times 10^{-5}$ | $7.51 \times 10^{-5}$ |
| 3 layers | $4.44 \times 10^{-4}$ | $1.09 \times 10^{-4}$ | $1.13 \times 10^{-4}$ |
| 4 layers | $2.14 \times 10^{-3}$ | $6.17 \times 10^{-4}$ | - |

## B. Performance of the regular CNN model

The performance of the regular CNN model, in which field knowledge is not encoded, is evaluated using the testing dataset. For each porous structure described by a 128×128 binary image, its effective diffusivity is computed by the trained model using ~$1.5 \times 10^8$ FLOPs of calculations. This cost is close to the cost of multiplying two 400×400 matrices and requires ~4 ms on a laptop (Intel CPU 6600, 3.3 GHz, 8 GB memory, single core). As a comparison, computing the effectivity diffusivity using LBM requires ~ 1 hour on the same computer. Therefore, the computational cost of the CNN model is roughly six orders of magnitude smaller than the LBM simulations.

Table 2. The mean square error and mean truncated relative error of the various CNN models

| | Regular CNN trained using the loss function Equ. 7 | Regular CNN trained using the loss function Equ. 10 | Porosity-informed CNN | Regular CNN with preprocessed image as input |
|---|---|---|---|---|
| Mean square error | $8.64 \times 10^{-4}$ | $7.01 \times 10^{-4}$ | $7.64 \times 10^{-4}$ | $6.92 \times 10^{-4}$ |
| Mean truncated relative error | 68.8% | 41.7% | 59.2% | 29.7% |
| $R^2$ | 0.9903 | 0.9921 | 0.9914 | 0.9912 |

The mean square error (MSE) of the effective diffusivity of porous structures in the entire testing dataset is found to be $8.64 \times 10^{-4}$ (see Table 2). Since the porous structures feature diverse pore shape/topology and their effective diffusivity spans ~$10^{-10}$ to 1.0, such a small MSE suggests that the CNN model performs well for the complex porous structure examined here. Figure 3a further compares the effective diffusivity predicted by the CNN model and computed by LBM codes, and we observe that the CNN model's overall performance is again very good. It is worthwhile to compare the performance of the CNN model with that of empirical correlations. The Bruggeman equation has been widely adopted in predicting the effective diffusivity of porous structure and is given by $D_e = \varepsilon/\tau$ ($\varepsilon$: porosity, $\tau$: tortuosity). The tortuosity of porous media is commonly modeled using $\tau = \varepsilon^{-\beta}$. Hence, the Bruggeman equation predicts

$$D_e = \varepsilon^{\beta+1} \tag{8}$$

The parameter $\beta$ depends on the structure of porous media, e.g., the connectivity of pores. $\beta$ is equal to 0.5 for porous media made of packed spheres and is otherwise an empirical parameter that is strongly correlated with the tortuosity of the porous media.[1,29] Here, $\beta$ is taken as 2.0 because it describes the scaling of $D_e$ at large porosity relative well. Figure 3a compares the prediction of the Bruggeman equation with the LBM prediction. When the true $D_e$ is larger than ~0.6, the Bruggeman equation performs as well as the CNN model. However, as the true $D_e$ becomes smaller, which is mostly caused by the emergence of more tortuous transport pathway in the porous media, the Bruggeman equation systematically overpredicts the effective diffusivity and performs worse than the CNN model. Therefore, the CNN model can better capture the transport properties of porous media with complex structure than the Bruggeman equation.

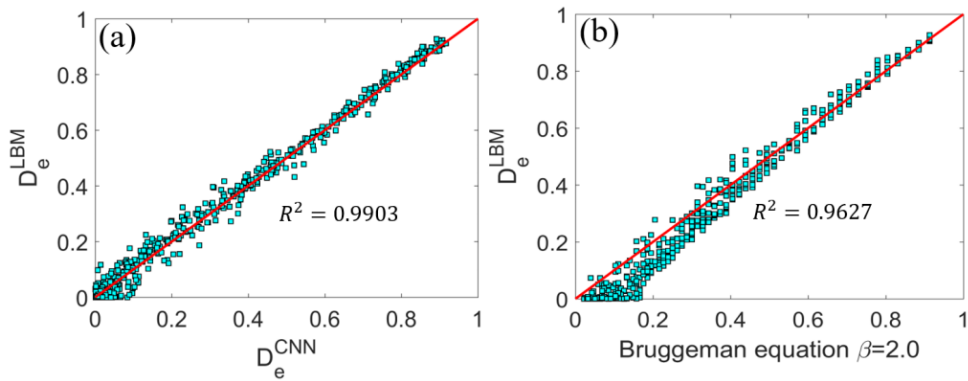

**Figure 3**. Effective diffusivity predicted by the CNN model (**a**) and the empirical Bruggeman equation (**b**).

We next quantify the predictive power of the CNN model *posteriori* systematically. Because the performance of the CNN model depends on the effective diffusivity of porous structure (see Fig. 4a), the porous structures in the testing dataset are divided into three groups: those with $D_e < 0.1$, those with $0.2 < D_e < 0.6$ and those with $D_e > 0.6$. The top row of Fig. 4 shows the distribution of the absolute error of $D_e$ predicted by the CNN model. In each group, over 95% of predictions have absolute error smaller than 0.1. Furthermore, the distribution of the absolute error is similar for all groups, which suggests that, during the training of the CNN model, the characteristics of all three groups are "learned" by the model.

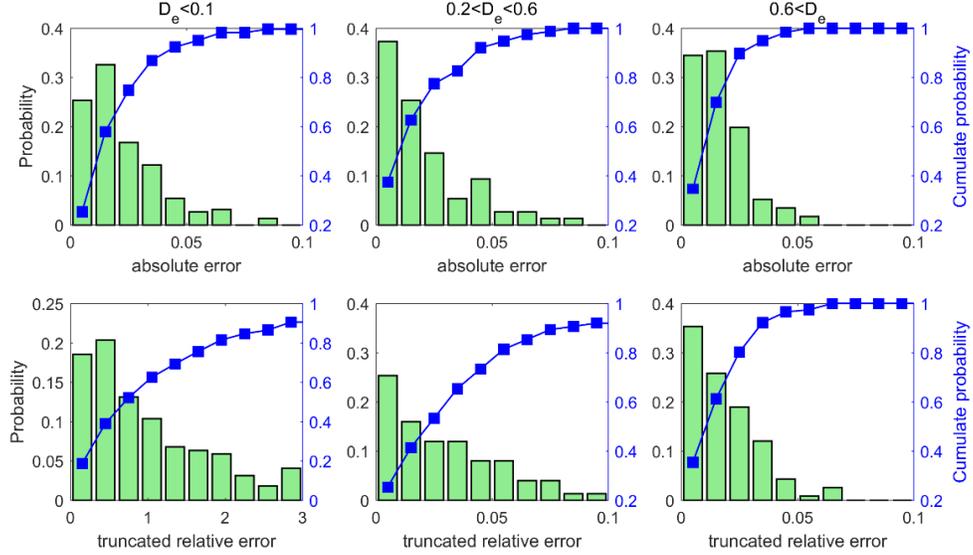

**Figure 4**. Distribution of the absolute error (top panels) and truncated relative error (lower panels) of the predictions of the CNN model for the porous structures in the testing dataset with $D_e < 0.1$, $0.2 < D_e < 0.6$ and $D_e > 0.6$. The CNN model is trained using the loss function based on the mean square error.

Another way to characterize the CNN model's predictive power is to study the relative error of its predictions. The standard calculation of relative errors, however, is not necessarily a good approach for the porous media studied here. Specifically, because many of our porous structures have extremely small $D_e$ (e.g., $10^{-10}$, see Fig. 1), a minute error in their $D_e$ can correspond to an enormous relative error even though such minute error has little practical impact on the prediction of transport in the porous media. While there is no unique way to address this issue, we define a truncated relative error, TRE, as

$$\text{TRE} = \begin{cases} \frac{|D_e^{\text{LBM}} - D_e^{\text{CNN}}|}{D_e^{\text{LBM}}} & \text{(if } D_e^{\text{LBM}} > D_e^{\text{thsd}}) \\ \frac{|D_e^{\text{LBM}} - D_e^{\text{CNN}}|}{D_e^{\text{thsd}}} & \text{(if } D_e^{\text{LBM}} < D_e^{\text{thsd}}) \end{cases} \qquad (9)$$

where $D_e^{\text{LBM}}$ and $D_e^{\text{CNN}}$ are the predictions by the LBM simulations (taken as ground truth here) and the CNN model, respectively. In this definition, only inaccuracy comparable to or larger than a threshold $D_e^{\text{thsd}}$ (taken as 0.01 hereafter to be specific) is thought to affect the practical application of the $D_e$ computed for the porous structure (e.g., in the prediction of pore clogging during the discharging of Li-air electrodes). Analysis of the predicted $D_e$ shows that the mean truncated relative error is 68.8% for the entire testing dataset (see Table 2). The bottom row of Fig. 4 further shows the distribution of the truncated relative error for the three groups of porous structure in the testing dataset. We observe that the truncated relative error is the larger in samples with smaller $D_e$: for samples with $0.2 < D_e < 0.6$, ~95% of the CNN predictions have a truncated relative error less than 10%; for samples with $D_e < 0.1$, ~60% of the CNN predictions have a truncated relative error less than 100%. These observations are consistent with the above observation that the magnitude of the absolute error similar among samples with different $D_e$.

The larger relative error in samples with small $D_e$ is expected because the regular CNN model is trained to minimize the MSE (i.e., a measure of the absolute rather than relative error) over the training set. To explore whether the CNN model can be tailored to give smaller relative errors (especially in samples with $D_e < 0.1$), we define a new loss function for training the CNN model as

$$L = 1/N \sum_{i=1}^{N}(TRE_i) \tag{10}$$

where $N$ is the number of porous samples in the training dataset, $TRE_i$ is the truncated relative error for each training data, which is defined in Eqs. (11). Training of the CNN model based on this loss function is able to converge. The CNN thus trained gives a truncated relative error of 41.7% when applied to the samples in the testing dataset, which is moderately better than that of the regular CNN model (68.6% see Table 2). Interestingly, the MSE of the new model (7.01×10⁻⁴) is also slightly smaller than that of the regular CNN model (8.64×10⁻⁴). This result indicates that the choice of the loss function during the training of a CNN model also affects how effectively the trained model performs, presumably because the tradeoff between how well the model is fitted for porous structures with low and high $D_e$ is shifted when different loss functions are used. The latter is evident when predictions by the newly trained CNN model is compared against the LBM predictions. As shown in Fig. 5a, compared to the regular CNN model, the new model performs better for samples with $D_e < 0.2$ but worse for samples with $D_e > 0.8$. These observations are corroborated by the distribution of the absolute errors in the three groups of samples in the testing dataset (see Fig. 5b, upper panels). The improved performance of the CNN model for samples with low $D_e$ is also evident in the distribution of the truncated relative error (see Fig. 5b, lower panels), e.g., for samples with $D_e < 0.1$, 85% of the predictions by the new CNN model is within a truncated relative error of 100%.

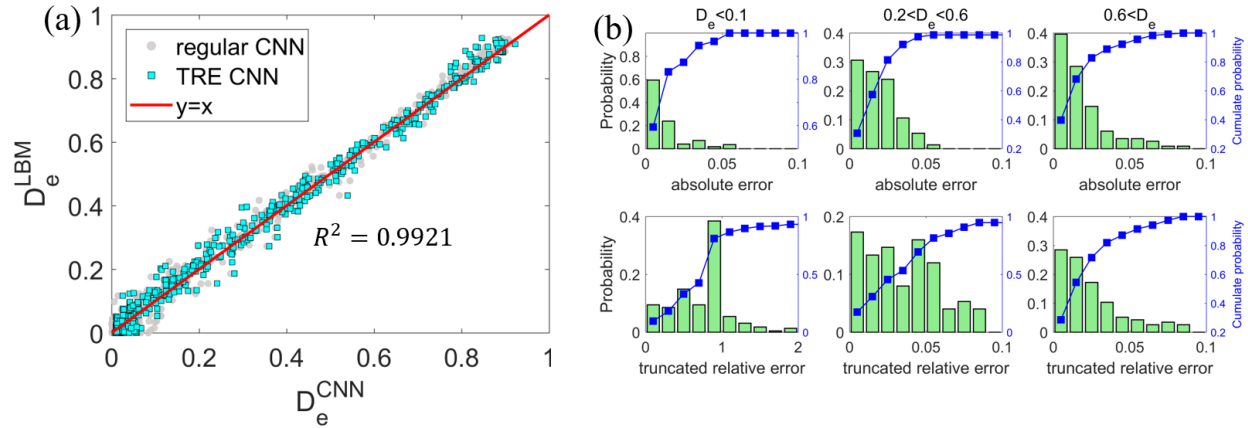

**Figure 5**. (**a**) The effective diffusivity predicted by the CNN model trained using a loss function based on the truncated relative error (Equation (9-10)). (**b**) Distribution of the absolute error (top panels) and truncated relative error (lower panels) of the predictions of the CNN model for the porous structures in the testing dataset with different $D_e$.

### C. Improving CNN models using field knowledge
#### C.1. Porosity-informed CNN model

In the previous section, we demonstrate that the CNN model performs well for porous structures with a wide range of porosity. Note that the key features determining the effective diffusivity of a porous structure are extracted through the convolutional layers and these features are mostly connected with the input images locally. Therefore, global features or features spanning large scale may not be effectively extracted using the CNN, which may compromise the predictive power of the CNN. Therefore, it may be useful to directly introduce physical parameters describing these features into the CNN model to improve its performance. As described in Section II, here we introduce the overall porosity of the porous media as an input to the first fully connected layer in the CNN. Although the complex porous structures spanning large scales are not easy to identify using CNN models, they can be easily identified based on knowledge of porous media.

The predictions of the porosity-informed CNN model are only marginally improved compared to the regular CNN, e.g., both the MSE and truncated relative error for the testing dataset are smaller than those of the regular CNN by about 12% (see Table 2). This is also evident in Fig. 6, where the CNN predictions for the entire testing dataset are shown. In particular, we observe that the limitations of the regular CNN model for porous structures with $D_e \lesssim 0.1$ are not greatly alleviated in the new model. The limited improvement of the new model may originate from several sources. First, porosity is not a reliable indicator of $D_e$, especially for porous structures with low $D_e$ (see Fig. 3b), because the transport is hindered also by complex features such as dead-end transport pathways and trapped space beyond the overall amount of pore space. Second, though porosity is added as additional neurons in the CNN, their impact may have already been captured in the feature maps generated by the filters in the convolutional layer. In this case, the input due to porosity is redundant and thus does not improve the performance of the CNN.

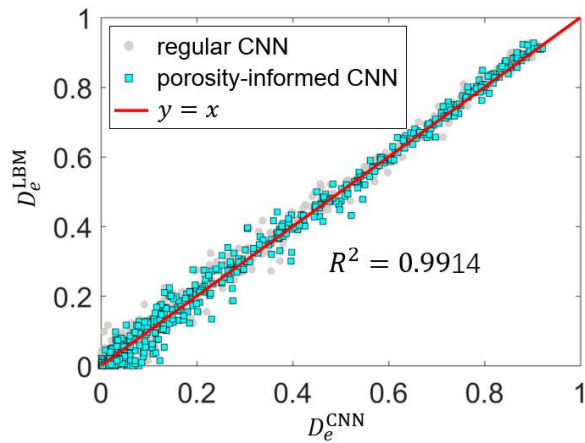

**Figure 6**. The effective diffusivity predicted by the porosity-informed CNN model.

### C.2. The CNN model using preprocessed porous structure as input

The CNN model illustrated in the previous sections works well for porous media with moderate to large diffusivity ($D_e > 0.2$) but exhibits relatively large error for porous samples with very small diffusivity (e.g., $D_e < 0.1$). This is closely related to the more complex transport behavior in porous media with very small diffusivities: in these media, the diffusion pathways are tortuous and there exist many trapped pores and dead-end paths. Since CNN models may not effectively extract features of these complicated structure spanning relatively large length scales using filters with small spatial extent, they do not perform well for such porous media. Nevertheless, these complex features are easily discerned based on field knowledge of porous media. Hence, here we explore the possibility of improving CNN prediction by processing the images of porous structures to remove dead-end and trapped pore spaces.

To improve the CNN prediction for porous media with small $D_e$, *all* datasets are optimized by removing the trapped and dead-end pore spaces. First, each binary image of the porous structures is labeled using 8-connected component labeling method[25,26] with periodic conditions on the top and bottom side (see Fig. 7a). This labeling method is consistent with LBM simulations based on the D2Q9 scheme, where each target position is associated with the nearby 8 directions. Next, in each labeled image, all components with labels across the inlet and outlet of the porous structure are kept, while other components (trapped pores and dead-end pores) are eliminated. Afterward, the porous structure is reconstructed with the remaining pore space as the effective pore space and the rest as the solid space. Figures 7a and 8b show a porous structure before and after the preprocessing described above. Clearly, although there exists large pore space in the regular porous structure, only part of it is available for molecular transport. Finally, the CNN model is trained using the preprocessed binary images and their corresponding $D_e$.

Using the CNN model trained above, we again calculated the effective diffusivity of the porous samples in the testing dataset from their preprocessed images (see Fig. 7c). Overall, the new predictions are in good agreement with the LBM predictions. As shown in Table 2, the MSE of the CNN predictions is 6.92×10$^{-4}$, which is 20% smaller than that of the regular CNN model. The mean truncated relative error is reduced from 68.8% in the regular CNN predictions to 29.7%. Examination of the distribution of the absolute and truncated errors of the predictions (see Fig. 7d) shows that the predictions are improved considerably for porous samples with $D_e < 0.6$, e.g., for samples with $D_e < 0.1$, over 70% of the predictions are within a truncated relative error of 20%. Therefore, leveraging field knowledge can help improve the predictions of the CNN model.

Even with the above improvement, for porous samples with $D_e < 0.1$, ~10% of the CNN predictions still have a truncated relative error >200%. Examination of these data points revealed that the true $D_e$ of the corresponding samples is very small ($\lesssim 10^{-4}$) but the CNN prediction are much higher. In these porous samples, only a few diffusion pathways contribute to the net diffusion flux. A representative case is shown in Fig. 7a-b. Although there is a large amount of pore space in the original porous sample (Fig. 7a), only a single diffusion path is found after the trapped pore spaces are identified (Fig. 7b). The net transport through these porous samples is thus dictated by a few narrowest throats along the long, tortuous pathway. Therefore, to accurately predict the effective transport property, both global feature (tortuous pathway) and local feature (critical throats) need be captured by the CNN model. This need can be difficult to meet by the convolutional layers, which retrieves relatively *local* characteristics of the porous structure. Indeed, similar situations are also frequently encountered in direct simulations of transport phenomena exhibiting multiple length and time scales (e.g., turbulent flows). Second, the number of samples having the above challenging geometrical features is limited and outweighed by other samples in the training dataset. Therefore, these features are likely not extracted accurately in the CNN model.

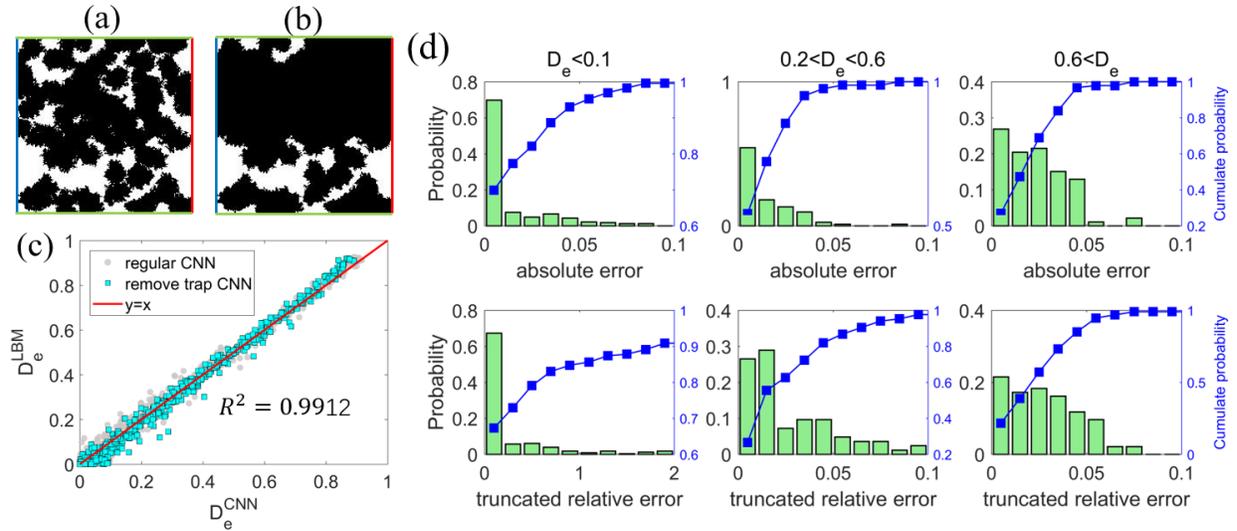

**Figure 7**. *Improving CNN prediction by preprocessing images of porous structures.* (**a-b**) Images of the regular (a) and processed (b) porous structure by removing the trapped pore space and dead-end transport pathways. The blue (red) lines on denote the inlet (outlet) of the molecule transport through the porous structure. The green lines denote the periodic boundaries. (**c-d**) The effective diffusivity predicted by the CNN model using the preprocessed image of porous structures (c) and the distribution of their absolute and truncated relative errors (d).

## IV. Conclusions

In summary, inspired by recent works on application of machine learning in porous media research, we develop deep learning models based on CNN for predicting the effective diffusivity of 2D porous media from their binary images. The computational cost of the model is six orders of magnitude lower than direct pore-scale calculations. The performance of the deep learning model is evaluated in detail by computing the mean square error, mean truncated relative error, and distribution of absolute and truncated relative errors for samples with different effective diffusivity. The effective diffusivity of realistic porous media spanning more than one order of magnitude ($0.1 \lesssim D_e < 1$) can be predicted well even when the porous media contain complex and diverse topologies and have large variation of porosity (0.28-0.98). The performance of the deep learning model, especially when applied to porous media with $D_e \lesssim 0.1$, can be improved by training the model using a loss function based on the truncated relative error or preprocessing images of porous media to remove the trapped pore space and dead-end transport pathways. Improvement due to incorporation of porosity of porous media into the CNN, however, is marginal.

Our results demonstrated that the transport properties of realistic porous media with diverse/complex topologies can be predicted with good accuracy and little computational cost. While only 2D porous structure is considered here, extension of the model to 3D situations is possible given that CNN can handle high dimensional data effectively. Indeed, the recent success of extending 2D CNN model to 3D for human pose and gesture recognition[30,31] and medical image process problems[32] suggests that CNN can be highly effectively in tackling 3D problems. Nevertheless, there may exists caveats that makes CNN less effective in 3D than 2D for the transport in complex media, and extensive tests must be performed to confirm the feasibility of CNN in predicting the transport properties of 3D porous media.

Although all datasets are generated computationally here for generic porous materials, these data can also be generated experimentally for specific porous materials, e.g., porous structures can be built from CT-scan and X-ray microtomography,[33-35] and effective transport properties such as diffusivity and thermal conductivity can be measured experimentally. The experimentally generated data can be used to inform the computational reconstruction of porous structures for specific classes of porous materials (e.g., shales or ceramic matrix composites) and validate predictions of pore-scale simulations, thereby enabling the creation of high-fidelity computational models of many classes of materials. The creation of these models and pore-scale simulations of them by high-performance computing will produce large, high-quality datasets tailored to different classes of porous materials. Deep learning models based on these datasets will enable fast and accurate prediction of the properties of these materials and benefit their applications. These potential extensions of the present work, along with the predictive power of the CNN models demonstrated here, suggest that deep learning can be a powerful new tool in the future research of porous media.

The analysis of the performance of our current model points to the origins of its current limitations and directions for future development. Our present model does not perform very well when the porous media has a very low effective diffusivity. Such a limitation may be addressed through two possible approaches: *multiscale feature extraction* and *encoding of advanced geometrical properties*. Because the transport in low effective diffusivity porous media is often dictated by a few narrow throats along a few (or even just one) tortuous pathway, the accurate prediction of transport by deep learning models requires the geometrical features at both global and local scale to be extracted effectively and properly weighed during training. Research on multiscale feature extraction can likely benefit from work on feature extractions from complex systems such as turbulent flows. Another way to simultaneously consider both the local and global structure of porous media is to directly encode information of such structure into the deep learning model, thereby bypassing the need to extract them using the convolutional layers. For example, given the importance of pore connectivity and pore size distribution in the transport in porous media, quantitative measures of these properties may be encoded into the deep learning model in the future. Identifying the best information to encode into deep learning models will benefit from the immense field knowledge on transport in porous media accumulated by the community over the past decades.

**Acknowledgements:** We thank the ARC at Virginia Tech for generous allocations of computer time.

**Author Contributions:** RQ, QK, and WQT conceived the project, HW and WZF designed the project and performed all simulations; HW and WZF analyzed the simulation results with input from everyone. HW led the writing of the manuscript and everyone participated in editing the manuscript.

**Additional Information**

**Competing interests**: The authors declare no competing interests.

**Data availability**: The datasets generated during and/or analysed during the current study are available from the corresponding author on reasonable request.